\documentclass[aps,prl,amsmath,twocolumn,showpacs,letterpaper]{revtex4}

\usepackage{graphicx}
\usepackage{psfrag}
\usepackage{color}

\begin{document}

\title{Probing Space-time Foam with Photons: Suppression of Observable Effects due to Uncertainty in Optical Paths}
\author{Yanbei Chen and Linqing Wen}
\affiliation{Max-Planck-Institut f\"ur Gravitationsphysik, Am M\"uhlenberg 1, 14476 Potsdam, Germany}
\date{May 27, 2006}

\begin{abstract}
It was recently proposed to use extra-galactic point sources to constrain space-time quantum fluctuations  in the universe.  In these proposals, the fundamental ``fuzziness'' of distance caused by space-time quantum fluctuations have been directly identified with fluctuations in optical paths. Phase-front corrugations deduced from these optical-path fluctuations are then applied to light from extragalactic point sources, and used to constrain various models of quantum gravity. In particular, the so-called random-walk model has been claimed to be ruled out by existing astrophysical observations from the Hubble Space Telescope.  However, when a photon propagates in three spatial dimensions, it does not follow a specific ray, but would rather sample a finite, three-dimensional region around that ray --- thereby averaging over space-time quantum fluctuations all through that region.  We use a simple, random-walk type model to demonstrate that, once the appropriate wave optics is applied, the averaging of neighboring space-time fluctuations will cause much less distortions on the phase front. In our model, the extra suppression factor due to diffraction is $l_{\rm P}/\lambda$, which is at least {19} orders of magnitude for astronomical observations.
 \end{abstract}
 
 \pacs{95.75.Kz, 03.67.Lx, 04.60.-m, 98.62.Gq}
\maketitle

It was recently proposed to use extra-galactic point sources to constrain the quantum fluctuation in space time~\cite{NCvD,LH,RTG,CNvD}. It was argued that, space-time fluctuations 
cause random phase shifts of photon, and that these fluctuations accumulate throughout the very long light propagation path from the point source to the earth, causing wavefront distortion from a perfect spherical shape upon arrival at the earth.  The manner these fluctuations accumulate depend on the specific model of quantum gravity, and in particular, it has been claimed that the {\it random walk} model could be ruled out by existing imaging data from the Hubble Space Telescope. 
In this paper, we point out what we think is a serious omission in the theory so far employed by the above authors, and argue that the random walk model, once given a closer look, cannot be ruled out at all by current or any forseeable observations of extra-galactic point sources.   

In Refs.~\cite{NCvD,LH,RTG,CNvD}, the authors assumed photons originating form the point source to undergo a random phase shift $\Delta \phi$ due to space-time fluctuations. According to various models of quantum gravity, they have 
\begin{equation}
\label{eqphi}
\Delta \phi  \sim 2\pi (l_{\rm P}/\lambda)^{\alpha} (L / \lambda)^{1-\alpha}
\end{equation}
 where $l_{\rm P}$ is the Planck length, $\lambda$ is the wavelength of the light, $L$ is the light propagation length, and $\alpha$ is a  parameter that depends on the model of quantum gravity.   In particular, $\alpha=1/2$ corresponds to the so-called {\it random-walk model}, which can be understood as having the speed of light fluctuating dramatically at the scale of $l_{\rm P}$.   The relation~\eqref{eqphi} stems from a more fundamental ansatz of quantum gravity, in which the uncertainty in distance measurement $\delta L$ over a distance $L$ is given by 
 \begin{equation}
\label{eql}
 \delta L \ge L^{1-\alpha} l_{\rm P}^{\alpha}\,,
 \end{equation}
 which was first derived by Ng and van Dam  and then discussed by others~\cite{CNvD,SciAm,GLM,MM}.  Note that Eq.~\eqref{eqphi} is related to Eq.~\eqref{eql} by  $\Delta \phi  \sim 2\pi (\delta L /\lambda)$.
  
While Eq.~\eqref{eql} could be regarded as fundamental to the quantum gravity theory, Eq.~\eqref{eqphi} cannot, because it also depends on how light propagates: when a photon travels through a space-time region, it does not follow only one particular ray, whose length is subject to the fundamental ``fuzziness'' prescribed by Eq.~\eqref{eql},  but instead, it would {\it simultaneously} sample {\it an ensemble} of many different neighboring rays, each of which having a {\it potentially different} realization of the fundamental length fluctuation; the actual path-length fluctuation must then be given by an {\it averaging} among these different length fluctuations.   This allows $\Delta\phi $ to go below $2\pi \delta L/\lambda$.  Moreover, because the linear size of the sampling region can be much bigger than $l_{\rm P}$, the correlation length of fundamental quantum fluctuations, this averaging can dramatically suppress the actual $\Delta \phi$ from  $2\pi \delta L/\lambda$, or Eq.~\eqref{eqphi}.

But we note that, if we lived in a space-time with one time dimension plus one spatial dimension, then there would be no uncertainty in optical paths, and there would not be any suppression factor from~\eqref{eqphi}.

{\it Model of space-time fluctuation.} In the rest of this paper, we will elaborate the above argument using a simple model of space-time foam.   We consider the light as a {\it scalar wave} traveling in  Minkowski space-time, filled with a medium with random (yet static) distribution of refractive index $n(x,y,z) \equiv 1+\epsilon(x,y,z)$.  We suppose that $\epsilon$ has the following  translational invariant  spatial auto-correlation function:
\begin{eqnarray}
\label{eqscart}
&&\langle \epsilon(x',y',z') \epsilon(x'',y'',z'')\rangle  \nonumber\\
&=& a^2 \Pi(x'-x'') \Pi(y'-y'')  \Pi(z'-z'')
\end{eqnarray}
where  ``$\langle \ldots \rangle$'' stands for ensemble average,  $a$ is of order unity, and 
\begin{equation}
\Pi(x)=\left\{
\begin{array}{ll}
1 & \displaystyle |x| \stackrel{<}{_\sim} l_{\rm P} \\
0 & |x| \gg l_{\rm P}
\end{array}
\right.
\end{equation}
In this toy model, the coordinate speed of light propagation fluctuates strongly, if we zoom in on a small region with size comparable to the Planck scale --- this simulates light propagation in a space-time with Planck-scale quantum fluctuations. [Note that we have ignored temporal fluctuations.]  In addition, light-speed fluctuations in regions separated by more than the Planck length are independently from each other. 

For a moment, if we took this mental picture of speed-of-light fluctuation, forgot about the wave nature of light, and assumed that each photon would follow a distinct coordinate path in $(t,x,y,z)$, or if we assumed a space-time with only one time dimension plus one spatial dimension,  we could also derive Eq.~\eqref{eqphi} with $\alpha=1/2$. In order to do so, we divide the distinct coordinate path into $N_L = L /l_{\rm P}$ intervals, each with length $l_{\rm P}$. The phase-shift fluctuation in each interval is $\delta \phi \sim  2\pi l_{\rm P}/\lambda $, while fluctuations in different intervals are independent from each other. In this way, the total phase shift  of the photon does a ``random walk'' while the light propagates.  At the end of propagation, we have a photon-phase fluctuation of  
\begin{equation}
\label{phinaive}
(\Delta\phi)_{\rm 1D} \sim \sqrt{N_L} \delta\phi \sim \sqrt{l_{\rm P} L /\lambda^2}\,,
\end{equation}
which is exactly Eq.~\eqref{eqphi}, with $\alpha=1/2$.


{\it The Wave Equation.} Returning to the wave picture, we first write down the wave equation:
\begin{equation}
-[1+2\epsilon(x,y,z)]\frac{\partial^2\Phi}{\partial t^2} + \nabla^2 \Phi =0\,.
\end{equation}
Since our refractive-index perturbation is static, we can expand the total wave into two monochromatic pieces, the primary wave, $\Phi_0(x,y,z) e^{-i\omega_0 t}$  and the secondary wave, $\psi(x,y,z) e^{-i\omega_0 t}$. At leading order in $\epsilon$, we have 
\begin{equation}
\label{perturbedwe}
\left(\nabla^2 +\omega_0^2 \right) \psi(x,y,z) = - 2\omega_0^2 \epsilon(x,y,z)  \Phi_0(x,y,z)\,.
\end{equation}
For a point source, we assume 
\begin{equation}
\Phi_0 (x,y,z) = \frac{e^{i\omega_0 r}}{4\pi r} \,,\quad r \equiv \sqrt{x^2+y^2+z^2}\,.
\end{equation}
At a distance $L$, the secondary wave $\psi$ must be compared with the primary wave at this point to give the amount of modulation it induces. We define
\begin{equation}
\alpha + i\phi \equiv 4\pi L \psi e^{-i\omega_o L}\,.
\end{equation}
with $\alpha,\phi \in \mathbf{R}$.  If $\alpha^2 +\phi^2 \ll 1$,  $\alpha$ is the relative amplitude modulation, and $\phi$ is the phase modulation in radians. We also define the {\it total modulation},
\begin{equation}
\label{xidef}
\xi  \equiv \sqrt{\langle \alpha^2 + \phi^2\rangle} = 4\pi  L  \sqrt{\langle \psi \psi^* \rangle} \,,
\end{equation}
which is larger than the standard deviations of both the amplitude and the phase modulation.

{\it Summing Over Paths.}  To arrive at the answer quickly, we use the Huygens-Fresnel-Kirchhoff scalar diffraction theory, which is equivalent to applying the outgoing Green Function, and obtain~\cite{jackson}:
\begin{eqnarray}
\label{green}
\psi(\mathbf{x})\!\! &=&\!\! \int_{|\mathbf{x}'|<L}  -2\omega_0^2 \epsilon(\mathbf{x}')\Phi_0(\mathbf{x}') \frac{e^{i\omega_0 |\mathbf{x} - \mathbf{x}'|}}{4\pi |\mathbf{x} - \mathbf{x}'|}d\mathbf{x}' \nonumber \\
\!\!&=&\!\!\int_{|\mathbf{x}'|<L} \frac{e^{i\omega_0 |\mathbf{x'}|}}{4\pi|\mathbf{x}'|} [-2\omega_0^2\epsilon(\mathbf{x}')]\frac{e^{i\omega_0 |\mathbf{x} - \mathbf{x}'|}}{4\pi |\mathbf{x} - \mathbf{x}'|}d\mathbf{x}'.\;
\end{eqnarray}
Note that we have only considered contributions from fluctuations at distances smaller than $L$ to the point source. The integral~\eqref{green} can be interpreted as a path integral --- over all paths that consist of two straight sections (each associated with a propagator), and a deflection in the middle due to interaction with refractive-index fluctuations (associated with a coupling coefficient).  Paths with more than one deflection do not have to be taken into account in our linear treatment.  

If we discretize the integration domain, a sphere with volume $\sim  L^3$, into cells with linear size $\sim l_{\rm P}$ and volume $v_{\rm P} \sim l_{\rm P}^3$, we will get a total of  $N_{\rm tot} \sim L^3/l_{\rm P}^3$ individual cells, each of has a statistically independent fluctuation in $\epsilon$ with variance $a^2$ [Cf.~Eq.~\eqref{eqscart}]. We will then estimate that Eq.~\eqref{green} would give
\begin{equation}
\sqrt{\langle |\psi |^2}\rangle \sim \frac{ 2\omega_0^2 a}{L^2} v_{\rm P} \sqrt{N_{\rm tot}} \sim a\omega_0^2 \sqrt{l_{\rm P}^3/L}\,.
\end{equation}
According to Eq.~\eqref{xidef} and comparing with Eq.~\eqref{phinaive}, we have 
\begin{equation}
\label{phiorder}
\Delta\phi \stackrel{<}{_\sim}\xi \sim a\omega_0^2 \sqrt{l_{\rm P}^3 L} \sim (\delta\phi)_{\rm 1D} (l_{\rm P}/\lambda)\,.
\end{equation}
There is an extra suppression factor of  $l_{\rm P}/\lambda$, which arises from the fact that in Eq.~\eqref{green}, the intermediate point $\mathbf{x}'$ of the optical path has the freedom to go away from the axis connecting the source point and the field point, and sample through $N_{\rm tot} \sim L^3/l_{\rm P}^3$ independent fluctuations, instead of only $N_L \sim L /l_{\rm P}$ in the one-dimensional treatment.

A more rigorous calculation using Eqs.~\eqref{eqscart}, \eqref{xidef} and \eqref{green} only gives a more precise numerical factor:
\begin{eqnarray}
\label{xikirch}
\xi &=& \sqrt{\frac{\pi}{8}}a\omega_0^2 \sqrt{l_{\rm P}^3 L}\,.
\end{eqnarray}

{\it Spatial-Scale Cut-off.} To  study fluctuations at different spatial scales, we solve the same problem by decomposing the secondary wave into modes:
\begin{equation}
\psi(r,\theta,\phi) = \sum_{\ell m} [\psi_{\ell m}(r) Y_{\ell  m}(\theta,\varphi)].
\end{equation}
Here $Y_{\ell m}(\theta,\varphi)$ are Spherical Harmonics. They describe angular variations at scales of $2\pi/\ell $; at a radius $r$, that corresponds to transverse length scales of $2\pi r/\ell $, or transverse spatial frequency of $\ell /(2\pi r)$.  The modal decomposition of Eq.~\eqref{perturbedwe} is
\begin{eqnarray}
\label{sphpsi}
&&\left[\frac{1}{r}\frac{\partial}{\partial r}\left(r\frac{\partial}{\partial r}\right)  + \omega_0^2 -\frac{\ell  (\ell  +1)}{r^2}\right]\psi_{\ell  m}(r)   \nonumber \\
&=&-\frac{ \omega_0^2 e^{i \omega_0 r}\epsilon_{\ell  m}(r)}{2\pi r}\,, \quad 
\end{eqnarray}
with\begin{equation}
\epsilon_{\ell  m}(r) \equiv \int_0^{2\pi} d\varphi \int_0^\pi{\sin\theta d\theta } \left[\epsilon(r,\theta,\varphi) Y^*_{\ell  m}(\theta,\varphi)\right]\,,
\end{equation}
which satisfies 
\begin{equation}
\label{epslm}
\langle \epsilon_{\ell  m}(r)\epsilon^*_{\ell 'm'}(r')\rangle =  a^2 \delta_{\ell  \ell'}\delta_{mm'} \delta(r-r') l_{\rm p}^3 /r^2\,. 
\end{equation}
Here we simply assumed 
\begin{equation}
\label{epssph}
\langle \epsilon(\mathbf{x})\epsilon(\mathbf{x}')\rangle  = a^2 l_{\rm P}^3 \delta^{(3)}(\mathbf{x}-\mathbf{x}')\,.
\end{equation}
The spatial spectrum corresponding to this correlation function is identical to that in Eq.~\eqref{eqscart} at low spatial frequencies, but continues to exist in orders  higher than $1/l_{\rm P}$. In principle, those modes may also add incoherently to our output fluctuations, but as we shall see, their contributions will be negligible.  Solving Eq.~\eqref{sphpsi}, assuming regularity at $r=0$ and outgoing condition at $r =L$, we obtain, 
\begin{equation}
\label{psimode}
\psi_{\ell m}(L) = - \frac{ \omega_0^3 L h_{\ell }^{(1)}(\omega_0 L)}{2\pi} \int_0^L dr [r j_{\ell }(\omega_0 r) e^{i\omega_0 r} \epsilon_{\ell m}(r)],
\end{equation}
where $j_\ell $ and $h^{(1)}_\ell $ are Spherical Bessel function and first-kind Spherical Hankel function\footnote{In order to obtain Eq.~\eqref{psimode}, we break the right-hand side of Eq.~\eqref{sphpsi} into an integral over $\delta$-functions: $\int dr' F(r)\delta(r-r')$. For each $\delta(r-r')$, its contribution must be of the form $A j_\ell (\omega_0 r)$ for $r<r'$ (due to regularity at origin) and $B h_\ell ^{(1)}(\omega_0 r)$ for $r>r'$ (due to outgoing condition at infinity). Then $A$ and $B$ can be solved by applying continuity in $\psi$, and junction condition in $\partial\psi/\partial r$.}.

\begin{figure}[t]
\psfrag{xx}{$\xi_\ell^2$}
\psfrag{ll}{$\ell  /(\omega_0 L)$}
\includegraphics[width=0.4\textwidth]{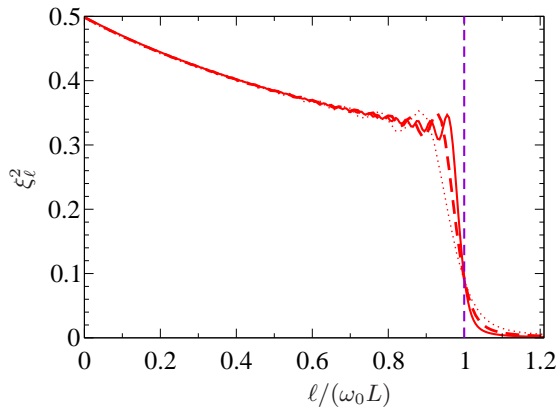}
\caption{Plots of $\xi_\ell^2$ as functions of $\ell  /(\omega_0 L)$, for cases with $\omega_0 L=50$ (dotted curve), 100 (dashed curve), and 200 (solid curve). All modes with $0<\ell/(\omega_0 L) <1.2$ are shown.
   \label{fig1} }
\end{figure}

From Eq.~\eqref{epslm} and~\eqref{psimode}, we obtain
\begin{eqnarray}
\label{psisphapp}
16\pi^2 L^2 \langle \psi_{\ell m} \psi_{\ell'm'}^*\rangle 
= \xi_{\ell m}^2 \delta_{\ell \ell'} \delta_{mm'}\,,
\end{eqnarray} 
with 
\begin{eqnarray}
\label{xil}
\xi_{\ell m}^2\equiv  4 a^2 \left|\omega_0 L h_{\ell }^{(1)}(L)\right|^2  (\omega_0 l_{\rm P})^3 \int_0^L j^2_\ell (\omega_0r) \omega_0 dr \,.
\end{eqnarray}
Note that $\xi_{\ell  m}$ is independent from $m$, which is a consequence of the rotation invariance of the refractive-index fluctuations. The total fluctuation at $r=L$  will then be
\begin{eqnarray}
\xi^2  &=&16\pi^2 L^2\langle \psi(L,\theta,\varphi)\psi^*(L,\theta,\varphi)\rangle \nonumber \\
&=& \sum_{\ell =0}^{+\infty}  \sum_{m=-\ell }^{+\ell } \xi_{\ell m}^2 Y_{\ell m}(\theta,\varphi)Y_{\ell m}^*(\theta,\varphi) \nonumber \\
&=& \sum_{\ell =0}^{+\infty} \frac{2\ell +1}{4\pi}\xi_{\ell 0}^2 \equiv \sum_{\ell =0}^{+\infty} \xi_\ell ^2\,.
\end{eqnarray}
Physically, $\xi_{\ell}^2$ describes fluctuations at the angular scale of $\sim 2\pi/{\ell}$, or transverse spatial scales of $2\pi L/\ell$, or transverse spatial frequency of $\ell /(2\pi L)$.  Inserting Eq.~\eqref{xil}, we have 
\begin{eqnarray}
\label{ximodal}
{\xi_\ell^2}&=&{a^2(\omega_0 l_{\rm P})^3}  \nonumber \\
&\times&   \frac{2 \ell +1}{\pi}\left| \omega_0 L h_\ell^{(1)}(\omega_0 L)\right|^2\int_0^{\omega_0 L} j_\ell^2(R)dR
\end{eqnarray}
We expect $\ell  \sim \omega_0 L$, or $2\pi L /\ell  \sim \lambda$  to be the turning point, because at this point the transverse spatial scale is comparable to the wavelength $\lambda$.

Mathematically, for $\ell   < \omega_0 L $, the Spherical Bessel and Hankel functions are wavelike at $r \sim L$, indicating {\it propagating waves}; for $\ell  > \omega_0 L$, the Spherical Bessel and Hankel functions are not wavelike at $r \sim L$, indicating {\it evanescent waves}.   In the limiting regimes of $\ell \ll \omega_0 L$  and $\ell \gg \omega_0 L$, $\xi_\ell$   can be evaluated analytically, using asymptotic expansions of Spherical Bessel functions:
\begin{equation}
\label{asymp}
\frac{\xi_\ell^2}{a^2 (\omega_0 l_{\rm P})^3} = \left\{ 
\begin{array}{ll}
1/2 \,, & \ell\ll \omega_0 L\,, \\
\omega_0 L /[(2\ell+1)^2\pi] \,, & \ell\gg \omega_0 L\,. 
\end{array}\right.
\end{equation}
Note that not only does $\xi_{\ell}^2$ approach $0$ at orders $\ell \gg \omega_0 L$, the summation of all these higher modes also gives a negligible contribution.  This qualitatively confirms a cut-off at the transverse scale of $\lambda$:  fluctuations at much finer scales do not generate secondary wave. This justifies our original use of Eq.~\eqref{epssph}.

In Fig.~\ref{fig1}, we study the transition zone of  $\ell \sim \omega_0 L$ numerically, for  moderately large values of $\omega_0 L =50$, 100 and 200, by plotting $\xi_{\ell}^2$ as a function of $\ell / (\omega_0 L)$. As $\omega_0 L \rightarrow +\infty$, $\xi^2_{\ell}$ asymptotes to a smooth, non-zero function for $\ell/(\omega_0 L)<1$, and to $0$ for $\ell/(\omega_0 L)>1$.  This means, in the realistic situation of $\omega_0 L \ll 1$,  $\ell/(\omega_0 L) =1$ is a sharp turning point between propagating and evanescent waves. 

It might seem difficult to evaluate the summation \eqref{ximodal} analytically. But since we are solving exactly the same problem as the previous section, it should be clear that [Cf.~Eq.~\eqref{xikirch}]
\begin{equation}
\xi = \sqrt{\sum_{\ell=0}^{+\infty} \xi_{\ell}^2}  = \sqrt{\frac{\pi}{8}} a\omega_0^2 \sqrt{ l_{\rm P}^3 L}  \sim (\Delta \phi)_{\rm 1D}(l_{\rm P}/\lambda)\,,
\end{equation}
as we have verified numerically in the cases $\omega_0 L =50$, 100 and 200. 

Having obtained the cut-off  length scale of $\lambda$, it is easy for us to offer a simple explanation to the extra suppression factor of $(l_{\rm P}/\lambda)$  using Fourier optics. In each transverse direction, the spatial spectrum of refractive-index fluctuation is flat, up to the spatial frequency of $1/l_{\rm P}$. Since only fluctuations with spatial frequencies below $1/\lambda$ propagates, we have a suppression of $l_{\rm P}/\lambda$ in ``power'', or $\sqrt{l_{\rm P}/\lambda}$ in linear amplitude. With two transverse directions, we then expect the suppression factor of $l_{\rm P}/\lambda$.

{\it Discussions and Summary.}  In this paper, we have calculated fluctuations on the phase front of an extra galactic point source, caused by Planck-Scale fluctuations in refractive index --- a toy model for space-time foam.  If diffraction of light were ignored, or if we assumed a space-time with one time dimension plus one spatial dimension,  our toy model  would give comparable results to previous esimates on the random-walk model~\cite{NCvD,LH,RTG,CNvD}. However, the diffraction of light forces us to average space-time fluctuations over all different possible optical paths that extend to all three spatial dimensions. This averaging filters out all fluctuations with transverse scales finer than the wavelength. In our model, this causes an extra suppression factor of $l_{\rm P}/\lambda$, with   [Cf.~Eq.~\eqref{eqphi}]
\begin{equation}
\label{xifinal}
\Delta\phi \stackrel{<}{_\sim}    \sqrt{l_{\rm P} L /\lambda^2}\,(l_{\rm P}/\lambda )
\end{equation}
This suppression is at least by 19 orders of magnitude for astronomical observations, if the wavelength of $\lambda=10^{-16}\,$m were to be used.  Numerically, we have
\begin{equation}
\Delta\phi \sim 10^{-26} \sqrt{{L}/{\,{\rm Gpc}}}\left({5\times10^{-7}\rm{\rm m}}/{\lambda}\right)^2,
\end{equation}
which unfortunately leaves us with no hope of detecting this random-walk model  with any conceivable astronomical observations. 

Even though in this paper we have only studied a static, random-walk-type model in which space-time fluctuations are local to each spatial region with size $l_{\rm P}$,  we believe the diffraction of light is a general enough phenomenon to affect significantly the use of photons to probe any other model of quantum gravity.  

This work is supported by the Alexander von Humboldt Foundation's Sofja Kovalevskaja Programme (funded by the German Federal Ministry of Education and Research).

\end{document}